\documentclass[useAMS]{mn2e}
\usepackage{amssymb,amsmath}
\usepackage{graphicx}

\title[Double-Peaked Emitter 3C390.3]
{Further Evidence for the Accretion Disk Origination of the Double-Peaked 
Broad H$\alpha$ of 3C390.3}
\author[Zhang X.-G.]
       {Xue-Guang Zhang$^{1,2}$\\
       $^1$Purple Mountain Observatory, Chinese Academy of Sciences,
             2 Beijing XiLu, NanJing, JiangSu, 210008, P. R. China \\
       $^2$Chinese Center for Antarctic Astronomy, NanJing, 
             JiangSu, 210008, P. R. China}

\date{}
\def\LaTeX{L\kern-.36em\raise.3ex\hbox{a}\kern-.15em
    T\kern-.1667em\lower.7ex\hbox{E}\kern-.125emX}

\begin{document}
\pagerange{\pageref{firstpage}--\pageref{lastpage}} \pubyear{2013}
\maketitle
\label{firstpage}

\begin{abstract}
    In the letter, under the widely accepted theoretical accretion disk 
model for the double-peaked emitter 3C390.3, the extended disk-like 
BLR can be well split into ten rings, and then the time lags between the 
lines from the rings and the continuum emission are estimated, based on 
the observed spectra around 1995. We can find one much strong correlation 
between the determined time lags (in unit of light-day) and the flux 
weighted radii (in unit of ${\rm R_G}$) of the rings, which is well 
consistent with the expected results through the theoretical accretion 
disk model. Moreover, through the strong correlation, the black hole masses 
of 3C390.3 are independently estimated as $\sim10^9{\rm M_{\odot}}$, the 
same as the reported black hole masses in the literature. The consistencies 
provide further evidence to strongly support the accretion disk origination 
of the double-peaked broad balmer lines of 3C390.3.
\end{abstract}

\begin{keywords}
Galaxies:Active -- Galaxies:nuclei -- Galaxies:Seyfert -- 
     quasars:Emission lines -- Galaxies:individual: 3C390.3
\end{keywords}

\section{Introduction}
    
   As one well-known double-peaked emitter and one well-studied mapped 
AGN, 3C390.3 has been studied for  more than four decades  
(Burbidge \& Burbidge 1971, Chen \& Halpern 1989, Dietrich et al. 1998, 
2012, Eracleous et al. 1995, Eracleous \& Halpern 1994, 2003, 
Flohic \& Eracleous 2008, Gezari et al. 2007, Gliozzi et al. 2006, 2011, 
Kollatschny \& Zetzl 2011, Leighly et al. 1997, Lewis \& Eracleous 2006, 
Netzer 1982, O'Brien et al. 1998, Perez et al. 1988, Popovic et al. 2011, 
Sambruna et al. 2009, Sergeev et al. 2002, 2011, Shapovalova et al. 2001, 
2010, Zhang 2011a, 2013, Zu et al. 2011). Based on the long-term 
variabilities of the double-peaked balmer lines and the continuum emission, 
the proposed double-stream model (Vellieux \& Zheng 1991, Zheng 1996) has 
been ruled out for 3C390.3 due to the un-obscured emitting region on the 
receding jet (Livio \& Xu 1997), and the binary black hole model 
(Begelman et al. 1980, Boroson \& Lauer 2009, Gaskell 1996, 
Zhang et al. 2007) has been ruled out for 3C390.3 due to the unreasonably 
large central BH masses (Eracleous et al. 1997), the proposed accretion 
disk model (see references listed above) has been widely accepted for 
3C390.3.

   Furthermore, besides the analysis for the double-peaked line profiles, 
based on the reverberation mapping technique (Blandford \& Mckee 1982, 
Horne et al. 2004, Peterson 1993), some simple information of the dominant 
gas motions in the BLR of 3C390.3 has been estimated through the response 
of different parts of the double-peaked broad emission lines, e.g., the 
line core, peaks and wings (Dietrich et al. 1998, 2012, Popovic et al. 2011, 
Sergeev et al. 2002, Shapovalova et al. 2001, 2010): the blue and red peaks 
of the double-peaked broad balmer lines vary with the same time delay 
relative to the continuum variations, which is well consistent with the 
expected results under the accretion disk model for the double-peaked 
broad lines. Furthermore, the results in the literature have shown that 
there are no time delays between line wings, peaks and line core of 
double-peaked broad balmer lines of 3C390.3.  

  However, besides the simple results above based on the reverberation 
mapping technique applied to the line wings, peaks and core, there is one 
another interesting way to check the proposed accretion disk model for 
3C390.3. Through the widely accepted accretion disk model, the proposed 
extended disk-like BLR of 3C390.3 can be well split into multiple rings 
with different radii. Apparently, time lag between one line from one of 
the rings and the continuum emission should sensitively depends on the 
radius of the ring. Moreover, the time lags are much different from the 
ones in the literature for the line wings, peaks and core, because the 
line core includes great contributions from the line photons coming from 
all the rings of the disk-like BLR of 3C390.3 (such as the following results  
shown in Figure 1).  Certainly, our results are much different from 
the results in Flohic \& Eracleous (2008): each observed line profile of 
3C390.3 are split into 36 parts, each part has one narrower wavelength range 
$\sim 20\AA$. However, in our letter, each observed line is split into ten 
parts, each part from the corresponding ring of the BLR has the full 
wavelength range of the observed H$\alpha$. To check the dependence of the 
time lags on the radii of the rings is our main objective of the letter, 
which could provide further evidence to support or against the accretion 
disk model for the double-peaked emitter 3C390.3. %This paper is organized 
%as follows. Section 2 gives our data procedures and main results. 
%Section 3 shows 
%our discussions and conclusions. 

\section{Main Results}
  
     As shown in our previous papers (Zhang 2011a, 2013, Paper I and 
Paper II), we have shown that the standard elliptical accretion disk 
model with few effects from probable bright spots and/or warped structures 
(Eracleous et al. 1995) can be well applied to describe the properties of 
the observed double-peaked broad H$\alpha$ of 3C390.3 observed around 1995. 
So that, in the letter, the standard elliptical accretion disk model 
is mainly considered, and there are no further considerations for the 
other models which can be well applied to 3C390.3 observed in different 
periods. And moreover, because the broad H$\alpha$ having stronger intensity 
and less contamination from other narrow emission lines, we mainly consider 
the 67 spectra with available broad H$\alpha$ from the AGNWATCH project 
(http://www.astronomy.ohio-state.edu/\~{}agnwatch/) (Dietrich et al. 1998).

    As we simply discussed in the Introduction, it is very 
interesting to check the time lags between the continuum emission and 
the lines from the rings with different radii, under the accretion disk 
model for 3C390.3. Here, the total proposed disk-like BLR of 3C390.3 
under the standard elliptical accretion disk model is evenly split   
into ten rings ($i=1,2,3...10$) with lower and higher boundaries, 
[$r_0 + \delta(r)\times i, r_0 + \delta(r)\times (i+1)$],
where $r_0$ and $r_1$ represent the inner and outer boundaries of the 
total BLR, $\delta(r)=(r_1-r_0)/10$ means the extended size of each ring. 
%$R_i$ means the radius of the $i$th ring. 
Based on the best fitted results for the 67 observed double-peaked broad 
H$\alpha$, the mean value of $r_0$ is around $\sim200{\rm R_G}$ and the mean 
value of $r_1$ is around $\sim1200{\rm R_G}$, where 
${\rm R_G = \frac{G\times M_{BH}}{c^2}}$ is the Schwarzschild radius. 
In other words, each observed double-peaked broad H$\alpha$ could be well 
split into ten parts (ten lines), and each part has apparent double-peaked 
line profile. Here, the BLR would not be split into more rings, otherwise 
light traveling time for each ring should be much shorter than the 
average time interval for the light curves of 3C390.3. After the 
consideration of the more recent black hole masses of 3C390.3 
$M_{BH}\sim10^{9}{\rm M_{\odot}}$ (Dietrich et al. 2012), the light 
traveling time for each ring with extended size of $\sim100{\rm R_G}$ 
is about 6days, which is one appropriate time value.  Figure 1 shows 
one example for the ten lines from the ten rings. It is clear that the 
line core defined in the literature includes apparent and strong 
contributions from the lines coming from all the rings. Therefore, to 
check the time lags between the continuum emission and the lines from the 
rings should be more meaningful. Then the intensities for the ten lines 
from the ten rings, $flux_i (i=1,2...,10)$, can be well determined by 
the accretion disk model. Certainly, we should note the values of 
$flux_i$ could not be directly used, before some procedures being applied 
to complete the flux calibration. 

%%%%%%%%%%%%%%%%%%%%%Figure 1%%%%%%%%%%%%%%%%%
\begin{figure}
\centering\includegraphics[height=6.5cm]{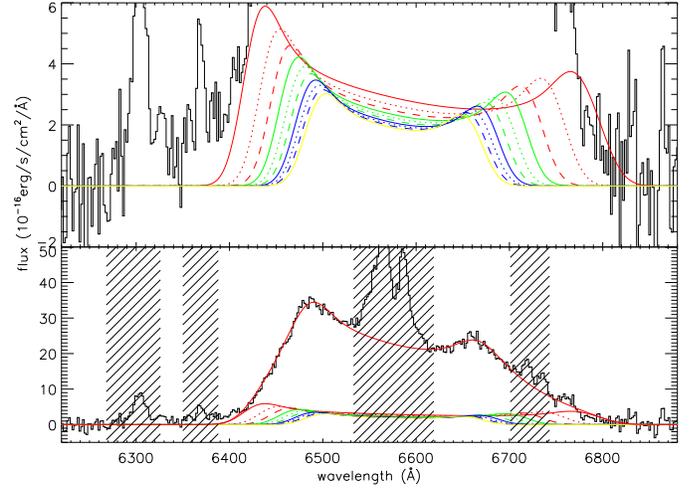}
\caption{Example for the ten lines coming from the ten rings of the 
disk-like BLR of 3C390.3. Top panel shows the clear profiles of 
the ten lines from the ten rings, different colors and line styles 
represent lines from different rings. Color of red, green, blue, yellow, 
and line style of solid line, dotted line and dashed line show the line 
from the ring with increasing radius. Bottom panel shows the sum of the 
ten lines (thick solid line in red color), which further represents the 
best fitted results for the observed double-peaked H$\alpha$ around 
JD-2449870. Moreover, the shadow areas in the bottom panel show the 
masked regions for the narrow lines.
}
\label{ring}
\end{figure}
%%%%%%%%%%%%%%%%%%%%%%%%%%%%%%%%%%%%%%%%%%%%%%

%%%%%%%%%%%%%%%Figure 2%%%%%%%%%%%%%%%%%%%%%
\begin{figure*}
\centering\includegraphics[width=17cm,height=17cm]{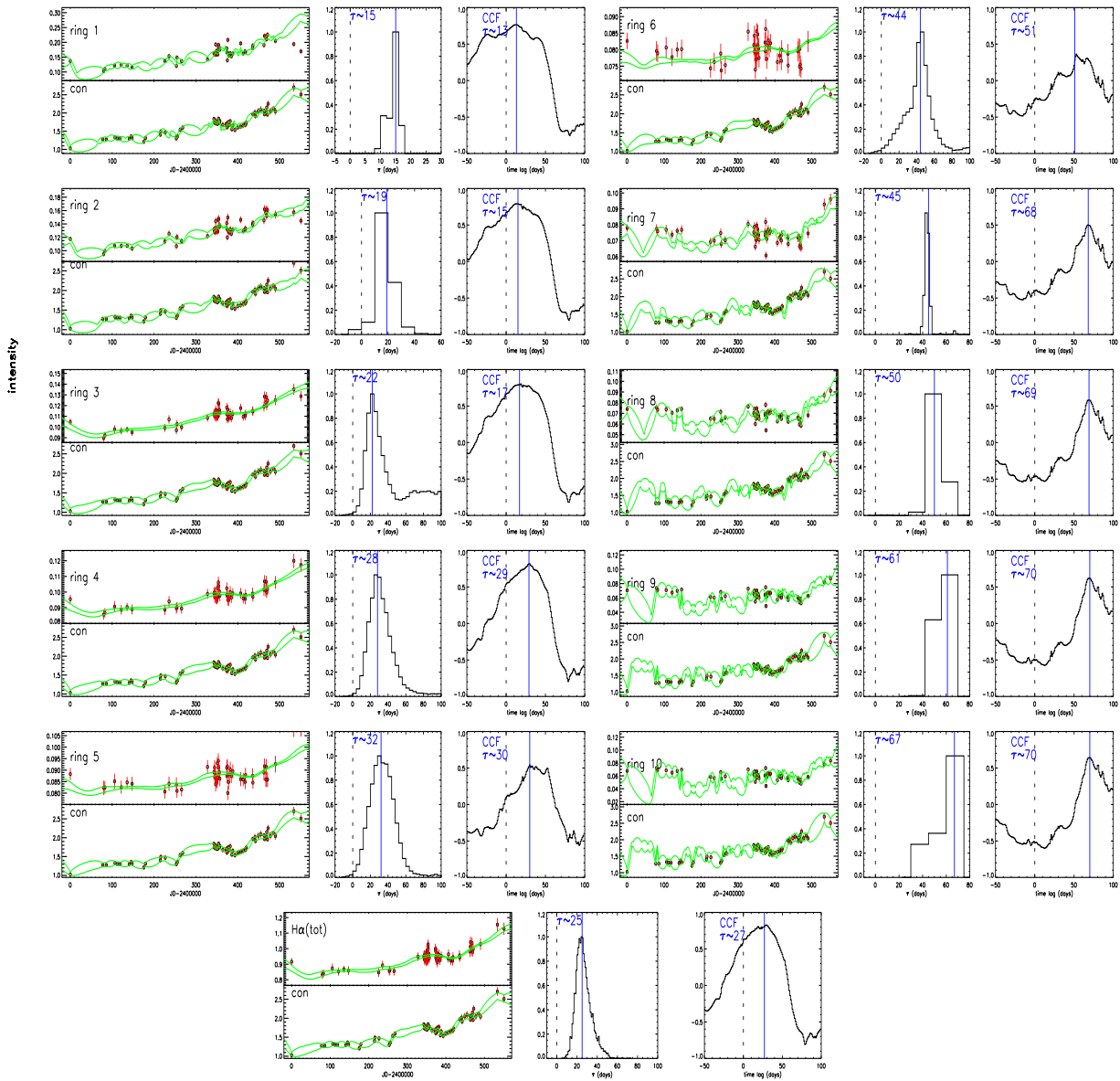}
\caption{The results based on the SPEAR method and the CCF method
applied to the light curves of the continuum emission and the ten lines 
from the ten rings. For each line and continuum, there are four panels, 
two panels show the light curves of the line and the continuum 
(red dots with error bars) and the the best descriptions for the light 
curves by the SPEAR method (green lines), one panel shows the distributions 
of the expected time lags between the line from the $i$th ring of BLR 
and the continuum through the SPEAR method (the distributions is normalized 
to have a maximum value equal to the peak value), and one panel shows the 
CCF results for the two light curves. The bottom panel shows the corresponding 
results for the continuum emission and the total observed broad H$\alpha$. 
In the figure, the vertical dotted 
lines show the positions with zero time lags, the blue solid lines 
show the peak positions of the distributions of the time lags through the 
SPEAR method, and the peak positions of the CCF results. The digital 
numbers in blue color gives the corresponding time lags. In the panels 
for the light curves, the unit for
the continuum is ${\rm 10^{-15}{\rm erg/s/cm^2/\AA}}$, and the unit for 
the emission lines is ${\rm 10^{-12}{\rm erg/s/cm^2}}$}
\label{results}
\end{figure*}
%%%%%%%%%%%%%%%%%%%%%%%%%%%%%%%%%%%%%%%%%%%%

   There are three steps being applied to complete the flux calibration for 
the line intensities of the ten lines from the ten rings. First and foremost, 
because the 67 spectra of 3C390.3 around 1995 are observed by different 
instruments in different observatories under different configurations, the 
measured line intensity based on the observed line profile should 
have low confidence level. Therefore, we collected the reliable values 
($F_{AGN}$ with corresponding uncertainties) listed in the AGNWATCH project, 
with the completed flux calibration having been applied (Dietrich et al. 1998) 
for the observed lines. Besides, the line 
intensities in the AGNWATCH project include contributions from the narrow 
lines around H$\alpha$. In order to ignore the effects of narrow lines on 
the following determined time lags, the intensities of the narrow lines 
should be subtracted from $F_{AGN}$. Here, based on the intensity of 
[OIII]$\lambda5007\AA$ in Dietrich et al. (1998) and the flux ratio 
of [NII] doublet and narrow H$\alpha$ to [OIII] in Dietrich et al. (2012), 
the total intensity of narrow lines around H$\alpha$ is determined as 
\begin{equation}
\begin{split} 
&F_{narrow}=flux(H\alpha_n) + flux([NII])\\
&\sim (3.53 + 1.34 + 0.44)\times flux([OIII])\\
&\sim117\times10^{-15}{\rm erg/s/cm^2}
\end{split}
\end{equation}
Finally, the available line intensities ($Flux_i (i=1,2,3,..,10)$) of the 
ten lines from the ten rings for each observed broad H$\alpha$ can be 
determined by the intensities ($flux_i$) from the direct observed 
lines without flux calibration, 
\begin{equation}
Flux_{i} = \frac{flux_i}{\sum flux_i}\times(F_{AGN}-F_{narrow})
\end{equation}
And, the uncertainty of $Flux_i$ can be determined by the 
uncertainties of $F_{AGN}$ and $flux([OIII])$. Then, the light curves of 
the continuum and the ten lines from the ten rings can be determined and 
shown in Figure 2. Here, the light curve of the continuum is the well-time 
binned one from the AGNWATCH Project 
(http://www.astronomy.ohio-state.edu/\~{}agnwatch/3c390/lcv/).  %Moreover, 
%if there were multiple observations in one night 
%(such as JD-2449868), then the weighted mean value was selected as the 
%value for the night. 

    Now, based on the light curves, we can check the dependence of 
the time lags for the ten lines on the radii of the ten rings. There are 
several commonly accepted CCF (Cross Correlation Function) methods to 
estimate time lag between two data series, such as the ZDCF method 
(Z-transfer Discrete Correlation Function, Edelson \& Krolik 1988, 
Peterson 1993, White \& Peterson 1994), the ICCF method (Interpolated 
Cross Correlation Function, Gaskell \& Peterson 1987, Peterson 1993), 
the MCCF method (Modified Cross Correlation Function, Koratkar \& Gaskell 
1989, Koptelova et al. 2006) etc.. More recently, Zu et al. (2011) have 
reported one new method to estimate time lag between continuum and emission 
line of AGN (SPEAR method, Stochastic Process Estimation for AGN
Reverberation), based on the assumption that all emission-line light curves 
are time-delayed, scaled, smoothed, and displaced versions of the 
continuum. This alternative approach fits the light curves directly using a 
damped random walk model (DRW model, Kelly et al. 2009, Kozlowski et al. 2010, 
MacLeod et al. 2010) and aligns them to recover the time lag and its 
statistical confidence limits. In the letter, both the ICCF method and 
the Zu's SPEAR method are applied to estimate the time lags between the 
continuum emission and the ten lines from the rings.

   When the ICCF method is applied, in order to ignore the effects 
of the large time gaps of the light curve as far as possible, only the 
data points within JD from JD-2449700 to JD-2450000 (around 300days) 
are considered. Then the corresponding uncertainty of the time lag can 
be determined by the Monte Carlo method as what we have done in 
Zhang et al. (2011b).  When the SPEAR method is applied, in order to 
obtain good enough results about the light curves through the DRW 
method, all the data points in the light curves are considered. Here, 
we do not describe the SPEAR method in detail, which can be found in 
Zu et al. (2011) and in the website: https://bitbucket.org/nye17/javelin. 
When the SPEAR method is applied, the number of walkers, burn-in 
iterations and sampling iterations for each walker are 300, 300 and 300 
respectively in the MCMC (Markov chain Monte Carlo method) analysis. 
Moreover, when the SPEAR method is applied, range from -10days to 100days 
is set as the boundaries for the time lag. The final results based 
on the ICCF method and the SPEAR method are shown in Figure 2.

  Based on the results shown in Figure 2 (especially the distributions 
of the time lags through the SPEAR method, and the CCF results), the time 
lags between the continuum emission and the ten lines from the ten rings 
are increasing with encreasing the radii of the rings. More clearer results 
are shown in Figure 3. In Figure 3, we show the correlations between the 
flux weighted radii of the ten rings (in unit of ${\rm R_G}$) and the time 
lags determined by the SPEAR method and the CCF method. Here, the flux 
weighted radii of the ten rings with lower and higher boundaries are 
calculated through the standard elliptical accretion disk model. For the 
$i$th ($i=1,2,3...10$) ring, there are 67 flux weighted radii based on the 
lower and higher boundaries for the 67 rings of the 67 observed broad 
H$\alpha$. Then, the mean value is accepted as the flux weighted radius of 
the $i$th ring, and the corresponding uncertainty is estimated as the value 
of the flux weight radius of the $i$th ring minus the minimum inner boundary 
for the 67 $i$th rings of the 67 observed spectra. It is clear that there 
are well consistent time lags determined by the SPEAR method and by the 
CCF method. Moreover, we can find the time lags between the ten lines and 
the continuum emission are strongly linearly correlated with the radii of 
the ten rings,
\begin{equation}
\begin{split}
\frac{R}{R_G} &= (19.45\pm2.15)\times\frac{\tau_{SPEAR}}{\rm light-days}\\
              &= (16.87\pm1.91\times\frac{\tau_{CCF}}{\rm light-days}
\end{split}
\end{equation}  
where $R$ means the flux weighted radii of the ten rings in unit of $R_G$. 
The corresponding values of Chi-square divided by number of degrees of 
freedom are 0.04 and 0.4 for results about $\tau_{SPEAR}$ and $\tau_{CCF}$. 
The results above are strongly support the accretion disk model
for 3C390.3.

\section{Discussions and Conclusions}

%%%%%%%%%%%%%%%%figure 3%%%%%%%%%%%%%%%%%%%%
\begin{figure}
\centering\includegraphics[height=6cm]{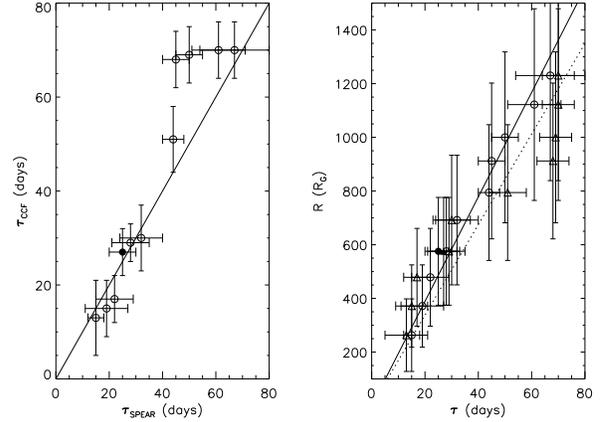}
\caption{On the correlation between the time lags determined by
the SPEAR method and by the CCF method (left panel), and the correlation
between the time lags and the radii of the rings (right panel). In the
left panel, the sold circle represents the values for the total observed
H$\alpha$. In the right panel, circles are for the time lags  based on
the SPEAR method, triangles are for values based on the CCF method, solid
circle and solid triangle are for the total observed H$\alpha$.
In the left panel, the solid line represents $\tau_{CCF}=\tau_{SPEAR}$.
In the right panel, the solid line and the dotted line are the best
fitted results (Equation 3) for the correlation about $\tau_{SPEAR}$
and for the correlation about $\tau_{CCF}$.
}
\label{check}
\end{figure}
%%%%%%%%%%%%%%%%%%%%%%%%%%%%%%%%%%%%%%%%%%%%

   Before further discussion, we firstly check whether our results about 
the time lags are reliable, especially for the time lag between the total 
observed broad H$\alpha$ and the continuum emission. In Zu et al. (2011), 
the time lag between the continuum emission and the H$\beta$ determined 
by the SPEAR method is about $27.9^{+2.4}_{-1.5}$ light-days for 3C390.3,  
which is well consistent with our result of $\sim25\pm5$ light-days 
between the continuum emission and the observed H$\alpha$ by the SPEAR 
method. Moreover, the time lag between the continuum and the observed 
H$\alpha$ is about $20\pm9$ light-days by the CCF method in 
Dietrich et al. (1998), which is consistent with our result $27\pm5$ 
light-days by the CCF method. The consistencies above indicate our 
procedures to estimate the time lags by the SPEAR method and by the 
CCF method are reliable. And therefore, the results shown in Figure 
2 and in Figure 3 have high confidence levels. In other words, the strong 
positive correlation shown in Figure 3 are reliable, which strongly support 
the accretion disk model for 3C390.3. 

    Besides to provide strong evidence for the accretion disk model 
for 3C390.3, the strong positive linear correlation shown in Figure 3 
could provide one another way to independently estimate the black hole 
masses of 3C390.3,
\begin{equation}
\frac{M_{BH}}{10^8{\rm M_{\odot}}} = \frac{1.}{0.005687\times\frac{R}{R_G}/\frac{\tau}{\rm light-days}}
\end{equation}
where the factor of '0.005687' is the value used to transfer the unit of 
${\rm R_G}$ to the physical unit of light-days.  Then, based on the 
Equation 3, the black hole masses of 3C390.3 are around 
$9.3^{+0.9}_{-1.1}\times10^8{\rm M_{\odot}}$ if 
$R\propto\tau_{SPEAR}$ being considered, and the black hole masses are 
around $10.4^{+1.3}_{-1.1}\times10^8{\rm M_{\odot}}$ if 
$R\propto\tau_{CCF}$ being considered. The black hole masses are very well 
consistent with the more recent result in Dietrich et al. (2012): 
$M_{BH}\sim10^9{\rm M_{\odot}}$. The results not only indicate our 
results in Figure 3 are reliable, but also provide one optional independent 
method to determine black hole masses for double-peaked AGN besides the 
M-sigma method (Ferrarese \& Merritt 2001, Gebhardt et al. 2000, 
G\"ultekin et al. 2009, Lewis \& Eracleous 2006) and the Virialization 
method (Dietrich et al. 2012, Peterson et al. 2004, Woo et al. 2010).

   Finally, we give our summary as follows.  The time lags between 
the ten lines coming from the well split ten rings of the disk-like BLR 
and the continuum emission have been calculated through the SPEAR method 
and the CCF method, under the theoretical accretion disk model. Then, 
we can find one strong correlation between the time lags and the radii of 
the rings. Moreover, based on the correlation, the independently determined 
black hole masses of 3C390.3 are well consistent with the more recent values 
in the literature.  The results above give further and strong  evidence 
for the accretion disk model for 3C390.3.

\section*{Acknowledgments}
Zhang X. G. gratefully acknowledge the anonymous
referee for giving us constructive comments and
suggestions to greatly improve our paper. ZXG gratefully
acknowledges the finance support from the Chinese grant
NSFC-11003043 and NSFC-11178003, and thanks the project of AGNWATCH
(http://www.astronomy.ohio-state.edu/\~{}agnwatch/) to make us
conveniently collect the spectra of 3C390.3.

\label{lastpage}
\end{document}